\documentclass{article}
\usepackage{spconf,amsmath,graphicx,bbm}
\usepackage{amssymb}
\usepackage{multirow,array,enumitem,adjustbox}
\usepackage{booktabs,threeparttable,makecell}
\usepackage[table,xcdraw]{xcolor}
\usepackage{xcolor}
\usepackage[colorlinks=true,citecolor=green]{hyperref}
\usepackage{cleveref}

\title{exploring binary classification loss for speaker verification}
\name{Bing Han, Zhengyang Chen, Yanmin Qian$^{\dagger}$  \thanks{$^\dagger$Corresponding Author. \\   \null\quad\ \ This work was supported in part by China NSFC projects under Grants 62122050 and 62071288, and in part by Shanghai Municipal Science and Technology Major Project under Grant 2021SHZDZX0102. Experiments have been carried out on the PI super-computer at Shanghai Jiao Tong University.
}}
\address{
  MoE Key Lab of Artificial Intelligence, AI Institute \\
  X-LANCE Lab, Department of Computer Science and Engineering \\
  Shanghai Jiao Tong University, Shanghai, China}

\begin{document}
%
\maketitle
\begin{abstract}
The mismatch between close-set training and open-set testing usually leads to significant performance degradation for speaker verification task. For existing loss functions, metric learning-based objectives depend strongly on searching effective pairs which might hinder further improvements. And popular multi-classification methods are usually observed with degradation when evaluated on unseen speakers. In this work, we introduce SphereFace2 framework which uses several binary classifiers to train the speaker model in a pair-wise manner instead of performing multi-classification. Benefiting from this learning paradigm, it can efficiently alleviate the gap between training and evaluation. Experiments conducted on Voxceleb show that the SphereFace2 outperforms other existing loss functions, especially on hard trials. Besides, large margin fine-tuning strategy is proven to be compatible with it for further improvements. Finally, SphereFace2 also shows its strong robustness to class-wise noisy labels which has the potential to be applied in the semi-supervised training scenario with inaccurate estimated pseudo labels. Codes are available in \url{https://github.com/Hunterhuan/sphereface2_speaker_verification}

\end{abstract}

\section{Introduction}
\label{sec:intro}
Speaker verification (SV) is the task of determining whether a pair of speech segments belong to the same speaker or not. Recently, with the thriving of deep neural networks (DNN), DNN-based speaker verification systems have obtained excellent performance when compared with traditional Gaussian Mixture Model (GMM)-based i-vector~\cite{dehak2010front}. Generally, a typical SV model consists of three parts: (1) a frame-level speaker feature extractor~\cite{snyder2018x, liu22h_interspeech}, (2) a pooling layer for statistic aggregation~\cite{wang2021revisiting, okabe2018attentive, zhao2022multi} and (3) a loss function for optimization.

For loss functions in SV, it can be mainly divided into two technical routes. Firstly, considering the open-set setting of SV task, it's reasonable to use contrastive learning-based metric objectives (eg. angular prototypical~\cite{chung2020defence}) to optimize the pair-wise similarity.
On the other hand, the softmax-based multi-class classifier is adopted to distinguish the different speakers in training set~\cite{xiang2019margin, liu2019large}. 
However, in verification task, both of them have some shortcomings. 
For metric learning-based methods, the performance strongly depends on the strategy to search effective pairs or triplets which is very time- and computation-consuming with the increasing of training samples number. For multi-classification methods, the embeddings produced by the DNN are not generalizable enough and performance degradation is observed when evaluated on unseen speakers due to the lack of similarity optimization explicitly~\cite{xiang2019margin}. 
Recently, several margin-based softmax variants~\cite{liu2017sphereface, deng2019arcface, wang2018additive, xiao2021adaptive, sun2020circle, ruida2022adaptive} are proposed to boost the discriminative power of speaker representation. They optimize the speaker embedding in a hyper-sphere space and encourage intra-class compactness by adding a margin to tighten the decision boundary. Although these multi-classification methods obtain significant performance gains, it's still difficult to ignore the mismatch between close-set training and open-set evaluation.

To alleviate the close-set assumption, in this paper,  we introduce a novel binary classification-based framework SphereFace2~\cite{wen2021sphereface2} for speaker verification. 
Unlike multi-classification training widely used before, it performs binary classification on hyper-sphere space which can effectively bridge the gap between training and evaluation, since both training and evaluation adopt pair-wise comparisons. 
Specifically, suppose there are $K$ speakers in the training set, it will construct $K$ independent binary classification objectives which regard data from the target speaker as positive samples and the others are negative. 
Experiments are conducted on Voxceleb~\cite{nagrani2017voxceleb, chung2018voxceleb2}, and the results illustrate that the SphereFace2 achieves better performance compared with the metric loss and multi-classification loss, which demonstrates that its pair-wise training manner can efficiently alleviate the mismatch between close-set training and open-set testing. Moreover, SphereFace2 also verifies its robustness against class-wise label noise, benefiting from the weak supervision of pair-wise labels.

\section{Related Work}
\subsection{Metric Learning-based Loss}
Prototypical loss~\cite{snell2017prototypical} is a widely used metric learning-based loss function in speaker verification. During the training, each mini-batch consists of a support set $S$ and a query set $Q$. In our implementation, we sample $N \times M$ utterances in each mini-batch, where $N$ is the speaker number and $M$ is the utterance number for each speaker. Besides, we consider the $M$-th utterance for each speaker as the query and the others as the support set.
Then the prototype $c_j$ for each speaker can be calculated by $c_j = \frac{1}{M-1}\sum^{M-1}_{m=1}x_{j,m}$. Then, $i$-th query can be classified against $N$ speakers based on softmax to optimize the distance between samples and prototypes:
\begin{equation}
L_{P} = -\frac{1}{N} \sum^{N}_{i=1}\log \frac{e^{S_{i,i}}}{\sum_{k=1}^{N}e^{S_{i,k}}}
\end{equation}
where $S_{i,k} = ||x_{i,M}-c_k||_2$, the squared Euclidean distance between the $i$-th query and $k$-th prototype. Besides, we can replace the L2-distance function with a cosine-based similarity metric to get angular prototypical loss~\cite{chung2020defence}:
\begin{equation}
    S_{i,k} = w\cdot \cos(x_{i,M},c_k)+b
\end{equation}
where $w$ and $b$ are learnable scale and bias parameters.


\subsection{Margin-based Softmax}
For margin-based softmax loss function, the general formula can be summarized as:
\begin{align}
    L_{M} = -\frac{1}{N} \sum^{N}_{i=1} \log \frac{e^{s\cdot \psi(\theta_{y_i})}}{e^{s\cdot \psi(\theta_{y_i})} + \sum_{j\ne y_i}e^{s\cdot \cos(\theta_j)}} 
\end{align}
where $\psi(\theta_{y_i})=\cos(m_1\theta_{y_i}+m_2)-m_3$ and $s$ is the scale factor to accelerate and stabilize the training. The $m_1$, $m_2$ and $m_3$ correspond to angular softmax (A-softmax)~\cite{liu2017sphereface}, additive angular softmax (AAM-softmax)~\cite{deng2019arcface} and additive margin softmax (AM-softmax)~\cite{wang2018additive} respectively. With these margins, the decision boundary is tightened which can explicitly 
enhance the similarity of intra-class samples and enlarge the distance between inter-class samples.

\section{SphereFace2: Binary classification}

Given $K$ speakers in the training set, SphereFace2 is designed to alleviate the mismatch between close-set training and open-set evaluation by explicitly constructing $K$ independent binary classification heads to perform the pair-wise comparison. Specifically, for the $i$-th sample in a batch, $\boldsymbol{x}_i\in \mathbb{R}^d$ represents the corresponding input to the classification layer and $y_i$ is the ground truth. For the projection head, we denote the weights of the $j$-th binary classifier by $\boldsymbol{W}_j$. Then the loss can be formulated as:
\begin{align}
\nonumber
    L_i = & \log(1+\exp(-\boldsymbol{W}_{y_i}^{\top}\boldsymbol{x}_i-b_{y_i})) \\ \nonumber
     & + \sum_{j\ne y_i}^K \log(1+\exp(\boldsymbol{W}_j^{\top}\boldsymbol{x}_i+b_j))
\end{align}
where $L_i$ is a summation of $K$ standard binary logistic regression losses. 
Following~\cite{liu2017sphereface, wang2018additive, deng2019arcface}, binary classification can also be optimized in hyper-sphere space by removing the bias $b_i$ and fixing all the binary classifier $||\boldsymbol{W}_j||_2=1$ and speaker embedding $||\boldsymbol{x}||_2=1$. Due to the lack of norm information, the variation range of cosine similarity is very small. So another parameter $s$ is introduced to scale the cosine similarity for accelerating and stabilizing the optimization~\cite{wang2018additive}:
\begin{align}
\nonumber
    L_i = & \log(1+\exp(-s \cdot \cos(\theta_{y_i})) \\ \nonumber
     & + \sum_{j\ne y_i}^K \log(1+\exp(s \cdot \cos(\theta_j))
\end{align}

For $K$ independent binary classifier and a sample $x_i$, they can only construct one positive sample and $K-1$ negative sample which is highly imbalanced. A simple but effective method is to introduce a weight parameter $\lambda \in [0,1]$ to balance the gradients for positive and negative samples. Then, the loss function becomes:
\begin{align}
\nonumber
    L_i = & \lambda \log(1+\exp(-s\cdot \cos(\theta_{y_i})) \\ \nonumber
     & + (1-\lambda) \sum_{j\ne y_i}^K \log(1+\exp(s\cdot \cos(\theta_j))
\end{align}

For multi-classification soft-based loss~\cite{liu2017sphereface, deng2019arcface, wang2018additive}, the decision boundary among different classes is not unified since there exists a competition among different classifiers and the boundary will be largely affected by the neighbor classifiers. As for SphereFace2, it can avoid such competition by utilizing $K$ independent binary classifiers, and then achieve a universal confidence threshold 0 ($s\cdot \cos(\theta_{y_i})=0$). However, it's difficult to achieve the universal threshold 0 in practice. Thus, the bias that was removed before comes back again to improve the training stability:
\begin{align}
\nonumber
    L_i = & \lambda \log(1+\exp(-s\cdot \cos(\theta_{y_i}) -b)) \\ \nonumber
     & + (1-\lambda) \sum_{j\ne y_i}^K \log(1+\exp(s\cdot \cos(\theta_j)+b))
\end{align}
where $b$ means the bias term. Then, the bias $b$ becomes the universal confidence threshold for all the binary classifiers and the decision boundary is turned into $s\cdot \cos(\theta_{y_i})+b=0$ which can increase the stability of training. 

The introduction of large margin penalty~\cite{liu2017sphereface,deng2019arcface,wang2018additive} on decision boundary, which can enforce the intra-class tightness and the inter-class discrepancy, has boosted the verification performance significantly. Similarly, an additive angular margin is added to SphereFace2 framework on two sides including positive and negative samples. Then the loss function can be formulated as:
\begin{align}
\nonumber
    L_i = & \lambda \log(1+\exp(-s\cdot (\cos(\theta_{y_i}) - m )-b)) \\ \nonumber
     & + (1-\lambda) \sum_{j\ne y_i}^K \log(1+\exp(s\cdot (\cos(\theta_j) + m)+b))
\end{align}
where $m$ is the adjustable margin parameter which can be used to further tighten the boundary. The final decision boundary of positive and negative samples are $s\cdot(cos(\theta_{y_i})-m)+b=0$ and $s\cdot(cos(\theta_{y_i})-m)+b=0$ respectively. 

A large inconsistency between the positive and negative pairs' score distribution is observed in~\cite{wen2021sphereface2}, and this discrepancy will make it difficult to find a threshold to distinguish the positive pairs due to the large overlap. To tackle this problem, a similarity adjustment method is proposed to map from angle to similarity score during training for discriminative distribution. Then, the final loss function can be summarized as:
\begin{align}
\label{equ:sphere_c}
    L_i = & \lambda \log(1+\exp(-s\cdot (g(\cos(\theta_{y_i})) - m )-b)) \\ \nonumber
     & + (1-\lambda) \sum_{j\ne y_i}^K \log(1+\exp(s\cdot (g(\cos(\theta_j)) + m)+b))
\end{align}
where $g(z) = 2(\frac{z+1}{2})^t-1$ is a mapping function to adjust the score similarity distribution.

It is noteworthy that all types of angular margins are compatible with SphereFace2. In addition to the Additive-type margin~\cite{wang2018additive} in Equation~\ref{equ:sphere_c}, we also explore the ArcFace-type margin~\cite{deng2019arcface} and a combination of two types margins which are denoted as SphereFace-A and SphereFace-M respectively.


\section{Experiments Setup}
\subsection{Dataset}
In our experiment, we trained all the systems on the development set of Voxceleb2~\cite{chung2018voxceleb2}, which contains 1,092,009 utterances among 5,994 speakers.  The evaluation trials in our experiment include three cleaned version trials Vox1-O, Vox1-E and Vox1-H constructed from 1251 speakers in Voxceleb1~\cite{nagrani2017voxceleb}. In addition, the validation trials from VoxSRC 2020 and VoxSRC 2021 are introduced to evaluate the performance on hard trials.

\subsection{Training Detail}
To explore the extreme performance of SphereFace2, three online data augmentation methods including adding noise~\cite{musan2015}, reverberation\footnote{\url{https://www.openslr.org/28}} and speed perturbation~\cite{thienpondt2020idlab} are applied here for robust training. The length of training samples is 2 seconds and we extract 80-dimensional Fbank with 25ms length Hamming windows and 10ms window shift as the input feature, while no voice activity detection (VAD) is involved here. The encoder we adopt is 32 channels ResNet34~\cite{zeinali2019but} with statistic pooling, and stochastic gradient descent (SGD) with momentum of 0.9 and weight decay of 1$e$-4 is employed as the optimizer to train the model. The whole training process will last 150 epochs and the learning rate decrease from 0.1 to 1$e$-5 exponentially. As for large margin fine-tuning \cite{thienpondt2021idlab}, the initial learning rate is set to 1$e$-4 and we train the models with only 5 epochs. It should be noted the margin is set to 0.35 and the segment duration increases to 6s in this stage.

\subsection{Evaluation Metrics}
For evaluation, we use cosine distance as the scoring criterion. After that, adaptive score
normalization (A-snorm)~\cite{cumani2011comparison} and quality-aware score calibration \cite{thienpondt2021idlab} applied for further improvements. Performance is measured in terms of the equal error rate (EER) and the minimum detection cost function (minDCF).

\section{Results and Analysis}

\begin{table*}[ht!]
\centering
\caption{\textbf{Voxceleb and VoxSRC results comparison between different loss functions.} LM-FT denotes the large margin fine-tuning strategy. For AM and AAM, the margin and scale are set to 0.2 and 32 respectively. And for circle loss, the margin and scale are 0.25 and 64 following the setup in~\cite{xiao2021adaptive}. For A-softmax, the margin is 4.}
\begin{adjustbox}{width=.99\textwidth,center}
\begin{threeparttable}
\begin{tabular}{lcccccccccc}
    \toprule 
       \multirow{2}{*}{Loss Function} & \multicolumn{2}{c}{Vox-O} & \multicolumn{2}{c}{Vox-E} & \multicolumn{2}{c}{Vox-H} & \multicolumn{2}{c}{VoxSRC20-val} & \multicolumn{2}{c}{VoxSRC21-val} \\
      \cline{2-11}
      & DCF$_{0.01}$ & EER & DCF$_{0.01}$ & EER  & DCF$_{0.01}$ & EER & DCF$_{0.05}$ & EER  & DCF$_{0.05}$ & EER \\
      \hline
      Angular Prototypical~\cite{chung2020defence} &  0.2286 & 1.356 & 0.1927 & 1.468 & 0.2885 & 2.699 & 0.2366 & 4.100 & 0.3361 & 6.614 \\ \hline
      Softmax &  0.1425 & 1.324 & 0.1532 & 1.292 & 0.2274 & 2.295 & 0.1955 & 3.549 & 0.2185 & 4.166 \\
      Circle Loss~\cite{sun2020circle, xiao2021adaptive} & 0.1014 & 0.946 & 0.1166 & 1.031 & 0.1702 & 1.823 & 0.1547 & 2.878 & 0.1743 & 3.147 \\
      A-Softmax~\cite{liu2017sphereface} & 0.1200 & 0.984 & 0.1300 & 1.087 & 0.1992 & 1.930 & 0.1628 & 3.003 & 0.2115 & 3.771 \\
      AM-Softmax~\cite{wang2018additive} &  0.0914 & \textbf{0.840} & 0.1147 & 0.987 & 0.1743 & 1.796 & 0.1553 & 2.919 & 0.1875 & 3.453 \\
      AAM-Softmax~\cite{deng2019arcface} &  0.0840 & 0.861 & 0.1122 & 0.996 & 0.1749 & 1.767 & 0.1531 & 2.830 & 0.1925 & 3.450 \\  
      \hline
      SphereFace2-A &  \textbf{0.0690} & 0.862 & 0.1069 & 0.993 & \textbf{0.1649} & 1.731 & 0.1494 & 2.761 & \textbf{0.1686} & \textbf{2.836} \\
      SphereFace2-M & 0.0851 & 0.914 & 0.1182 & 1.059 & 0.1772 & 1.831 & 0.1576 & 2.901 & 0.1838 & 3.060 \\ 
      SphereFace2 &  0.0757 & 0.877 & \textbf{0.1065} & \textbf{0.969} & 0.1699 & \textbf{1.726} & \textbf{0.1476} & \textbf{2.731} & 0.1741 & 3.067 \\ 
      \hspace{2pt} + LM-FT & 0.0571 & 0.670 & 0.0852 & 0.809 & 0.1384 & 1.424 & 0.1242 & 2.345 & 0.1376 & 2.362 \\
    \hline \bottomrule
\end{tabular}
\end{threeparttable}
\label{table:res_diff_loss_function}
\end{adjustbox}
\vspace{-0.5cm}
\end{table*}

\subsection{Comparison between Different Loss Functions}
\label{ssec:loss_res_compare}
In this section, we first give a results comparison between different loss functions in Table.~\ref{table:res_diff_loss_function}. In~\cite{chung2020defence}, metric learning-based objectives are boosted and achieved competitive performance with the classification-based losses when there is no data augmentation. 
However, the metric learning-based objectives require large batch-size to mine enough negative pairs and are sensitive to the hard negative mining strategy. Besides, equipped with more extensive data augmentation and advanced training strategies, the results in Table~.\ref{table:res_diff_loss_function} show that the classification-based losses have a more obvious advantage over the angular prototypical loss.

Among all the multi-classification losses, it's obvious to find a performance leap when the margin penalty is introduced to boost the discriminative power of speaker representation compared with traditional Softmax. 
In addition, SphereFace2 replaces the multi-classification with $K$ binary classifiers, and trains the model in pair-wise learning paradigm. In Table.~\ref{table:res_diff_loss_function}, SphereFace2 based loss functions surpass all other loss functions including metric- and softmax-based objectives, especially in hard trials (VoxSRC20-val and VoxSRC21-val). Moreover, the results of SphereFace2-A and SphereFace-M are also provided in Table.~\ref{table:res_diff_loss_function}, and we observe that different types of angular margins perform well and obtain similar performance.

Large margin fine-tuning (LM-FT)~\cite{thienpondt2021idlab} is a training strategy to further optimize the inter- and intro-class distance by enlarging the margin and duration, which is widely used in building challenge systems. And we find that LM-FT is also compatible with SphereFace2 and leads to a great performance gain.

Finally, we provide an ablation study to analysis the effect of hyperparameters $\lambda$, $t$, $s$ and $m$ in SphereFace2 loss, and the results are listed in Table~\ref{table:res_ablation}. According to the results, we observe that the SphereFace2 achieve the best performance under $\lambda=0.7$, $t=3$, $s=32$ and $m=0.2$.

\begin{table}[ht!]
\centering
\caption{\textbf{Ablation study of hyperparameters $\lambda$, $t$, $s$ and $m$.} Results are given with EER(\%).}
\begin{adjustbox}{width=.49\textwidth,center}
\begin{threeparttable}
\begin{tabular}{ccccccccc}
    \toprule 
      $\lambda$ & $t$ & $s$ & $m$ & Vox-O & Vox-E & Vox-H & VoxSRC20 & VoxSRC21 \\ \hline
      \cellcolor[HTML]{EFEFEF}0.7 & 3 & 32 & 0.2 & 0.877 & \textbf{0.969} & \textbf{1.726} & \textbf{2.731} & \textbf{3.067} \\
      \cellcolor[HTML]{EFEFEF}0.8 & 3 & 32 & 0.2 & \textbf{0.835} & 0.976 & 1.728 & 2.780 & 3.083 \\ \hline
      0.7 & \cellcolor[HTML]{EFEFEF}2 & 32 & 0.2 & 0.808 & 0.989 & \textbf{1.724} & 2.821 & 3.183 \\
      0.7 & \cellcolor[HTML]{EFEFEF}3 & 32 & 0.2 & 0.877 & \textbf{0.969} & 1.726 & \textbf{2.731} & \textbf{3.067} \\
      0.7 & \cellcolor[HTML]{EFEFEF}4 & 32 & 0.2 & \textbf{0.829} & 0.980 & 1.750 & 2.834 & 3.160 \\ \hline
      0.7 & 3 & \cellcolor[HTML]{EFEFEF}24 & 0.2 & \textbf{0.761} & 1.000 & 1.800 & 2.849 & 3.407 \\
      0.7 & 3 & \cellcolor[HTML]{EFEFEF}32 & 0.2 & 0.877 & \textbf{0.969} & \textbf{1.726} & \textbf{2.731} & \textbf{3.067} \\
      0.7 & 3 & \cellcolor[HTML]{EFEFEF}40$^\dagger$ & 0.2 & - & - & - & - & - \\ \hline
      0.7 & 3 & 32 & \cellcolor[HTML]{EFEFEF}0.1 & \textbf{0.824} & 1.000 & 1.806 & 2.878 & 3.333 \\
      0.7 & 3 & 32 & \cellcolor[HTML]{EFEFEF}0.2 & 0.877 & \textbf{0.969} & \textbf{1.726} & \textbf{2.731} & \textbf{3.067} \\
      0.7 & 3 & 32 & \cellcolor[HTML]{EFEFEF}0.3 & 0.909 & 1.056 & 1.838 & 2.947 & 3.263 \\
      \hline
    \bottomrule
\end{tabular}
\begin{tablenotes}
\item $^\dagger$: for parameter $s$, 40 is too large to train.
\end{tablenotes}
\end{threeparttable}
\label{table:res_ablation}
\end{adjustbox}
\end{table}

\subsection{Robustness on Noisy Label}
All the loss functions we discussed in section \ref{ssec:loss_res_compare} are supervised training loss, which require precisely labeled data. Although the Voxceleb dataset is collected through an automated pipeline \cite{nagrani2017voxceleb,chung2018voxceleb2}, the authors found very few label errors after manual inspection \cite{chung2018voxceleb2}. Thus, it is curious how well these algorithms perform on data with noisy labels. In this section, we select the best performed multi-classification loss AAM-softmax and compare it with the SphereFace2 loss. In this experiment, we randomly select 30\% of the data and change their labels before training.  The corresponding results are shown in Table \ref{table:res_noisy}. 
From the results, we find that the performance of the models trained on data with noisy labels has a clear drop in performance. Surprisingly, the performance degradation of the model trained with SphereFace2 loss is much smaller than the model trained with AAM-softmax. This is because the multi-class softmax function has very strong supervision, which pushes all the similarities between embedding and non-target centers smaller than the similarity between embedding and target centers. However, this strong supervision loss can be counterproductive when there are some data with noisy labels. In comparison, SphereFace2 loss has weaker supervision and is more robust to the noisy labels.

\begin{table}[ht!]
\centering
\caption{\textbf{EER(\%) results of different loss functions trained on data with noisy labels.} It should be noted that A-snorm and score calibration are not used here because these strategies require accurate speaker labels.}
\begin{adjustbox}{width=.45\textwidth,center}
\begin{threeparttable}
\begin{tabular}{l|cccc}
    \toprule 
        Loss Function& \multicolumn{2}{c}{AAM-softmax} & \multicolumn{2}{c}{SphereFace2} \\ \hline
        Noisy Proportion(\%) & 0 & 30 & 0 & 30 \\
      \hline
      Vox-O & 1.058 & 2.005 & 1.032 & 1.638 \\
      Vox-E & 1.147 & 2.106 & 1.060 & 1.665 \\
      Vox-H & 2.087 & 3.744 & 1.907 & 2.931 \\
      VoxSRC20-val & 3.398 & 5.669 & 3.120 & 4.646 \\
      VoxSRC21-val & 4.074 & 5.743 & 3.373 & 4.798 \\ \hline
    \bottomrule
\end{tabular}
\end{threeparttable}
\label{table:res_noisy}
\end{adjustbox}
\vspace{-0.5cm}
\end{table}


\section{Conclusion}
In this paper, we introduce SphereFace2, a binary classification-based loss function for speaker verification to alleviate the mismatch between open-set training and close-set testing. Experiments conducted on Voxceleb show its leading performance compared with popular metric learning or multi-classification based loss functions. Moreover, the large margin fine-tuning strategy is applicable to further boost the performance. Finally, SphereFace2 also shows its strong robustness to class-wise noisy labels which has the potential to be applied in the semi-supervised training scenario with inaccurate estimated pseudo labels.

\bibliographystyle{IEEEbib}
\bibliography{strings,refs}

\end{document}